October 3, 2015, Ver. 1.1

# 5G Ultra-Reliable Vehicular Communication

Erik G. Ström, Petar Popovski, Joachim Sachs

## Abstract

Applications enabled by Cooperative Intelligent Transport Systems (C-ITS) represent a major step towards making the road transport system safer and more efficient (green), and thus suited for a sustainable future. Wireless communication between vehicles and road infrastructure is an enabler for high-performance C-ITS applications. State-of-the-art communication systems for supporting low-latency C-ITS applications are based on IEEE 802.11 medium access control (MAC) and physical (PHY) layers. In this paper, we argue that a well-designed 5G system can complement or even replace these systems. We will review the C-ITS application requirements and explain how these are well aligned with the foreseen generic 5G service of ultra-reliable machine-type communication (uMTC). Key technology components suitable for constructing the uMTC service are identified: reliable service composition (RSC) and device-to-device (D2D) links for all-to-all broadcast communication, operational at high mobility and with varying degree of network assistance. Important problems for future studies, including radio-resource management, medium access control, and physical layer challenges, are discussed.

## 1 Introduction

History has shown that a new generation wireless system is introduced approximately each decade. Hence, we can expect that 5G will be gradually introduced on the market within a few years. Although it is not clear today exactly what 5G will be, some main trends can be discerned. For sure, mobile broadband will be at the heart of 5G, but 5G will be more than an enhanced 4G system. In fact, the METIS project, co-funded by the European Commission as an Integrated Project under the Seventh Framework Programme for research and development (FP7), has identified efficient support of machine-type communication (MTC) applications as an important feature of 5G.

In this paper, we will elaborate on a special type of MTC service called ultra-reliable MTC and how this service can be used to support very demanding C-ITS applications. The paper is organized as follows. Section 2 explains the distinguishing features of MTC compared to human-centric communication. Furthermore, MTC is subdivided into two main service classes: massive MTC (mMTC) and ultra-reliable MTC (uMTC). Section 3 defines important properties of the uMTC service: reliability, availability, and failure. Armed with sufficiently precise notions of these properties, we explain in Section 4 how the interaction between the applications that use the uMTC service and the system that provide the service can be orchestrated to tradeoff between service reliability and availability. Section 5 provides the necessary context for understanding the C-ITS application requirements. These requirements are then used in Section 6 to argue for how a matching uMTC service can be constructed from a number of key technology components. The paper is concluded in Section 7, along with an outline of the topics for future research.

## 2   Ultra-Reliable MTC and Massive MTC

The focus of cellular communication has been historically on human-centric communication, which includes, e.g., telephony, or providing some form of information to people from network-hosted servers, such as mobile Internet services or video streaming. In contrast, Machine-Type Communication (MTC) denotes the broad area of wireless communication with sensors, actuators, physical objects, embedded controllers and other devices not directly operated by humans. MTC is starting to play an increasing role in mobile networks and efforts have been put in the latest LTE releases to address MTC requirements [Sachs-15]. For 5G, efficient support for a wide range of different MTC use cases is considered as a key design target from the very beginning [Dahlman-14].

In [METIS D6.6], the METIS project describes three generic 5G services, extreme mobile broadband (xMBB), massive machine-type communication (mMTC), and ultra-reliable machine-type communication (uMTC).

mMTC refers to services where a typically large number of sensors monitor certain events or some kind of system state, which can be complemented with a form of actuation to control an environment.  mMTC plays a role in a wide range of areas, like, e.g., smart agriculture, smart city monitoring and operation, or asset tracking and logistics. Data transfers per device are typically infrequent and with relaxed delay requirements. At the same time simple, scalable and energy efficient communication is needed, which supports concentrations of massive numbers of devices in some areas, where devices need to remain very simple and can operate on batteries for many years.

In contrast, uMTC refers to services that provide very high reliability and often very short latencies. Hence, the uMTC service is suitable also for safety critical or mission critical applications, for which a service failure would have severe consequences. It is relevant for real-time control in automated cyber-physical systems, such as industrial process control, and it also appears as the key enabler of reliable communication for vehicles.

As described in more detail in Section 5, the state-of-the-art system for vehicular communication use dedicated spectrum, but with a medium access method that does not provide any service guarantees. This motivates the study how a 5G uMTC service can be provided for vehicular communication, as would support applications with unprecedented hard requirements on reliability. Indeed, this is one of the main topics of this paper.

## 3   uMTC Service Reliability, Availability, and Failure

The objective with this section is to make the notion of uMTC *service reliability*, *service availability*, and *service failure* more precise and illustrate some relations and tradeoffs between these.

In the following, we make a distinction between the system that provides the uMTC service (i.e., the "communication system") and the application that use the uMTC service (i.e., the "application"). The communication system consists of a number of protocol layers, that always include the medium access control (MAC) and physical (PHY) layers, but can also include network and other layers as needed (e.g., for multihop functionality). See Figure 1 for a simple illustration.

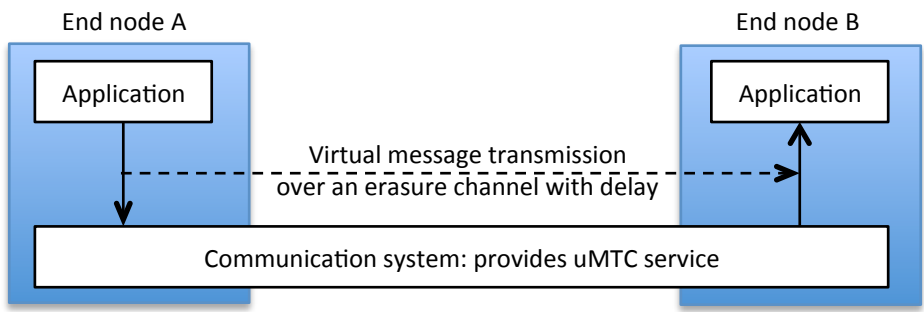

Figure 1: Illustration of two end-node applications using the uMTC service that is provided by the communication system. Perfect error detection is assumed, i.e., only error-free packets are delivered to the receiving end-node application. Hence, from the application point of view, messages are delayed and potentially lost in transmission.

Most, if not all, applications rely on timely delivery of data. Hence, the end-to-end latency, i.e., the time from when the transmitter-end application request to send a message until it is successfully delivered to the receiving-end application, must be limited. The application need of "timeliness" is easiest to model with hard deadlines, i.e., we consider messages whose end-to-end latency that exceeds the deadline as useless. It is noted that this is in contrast with soft deadlines, i.e., when the value of the message decreases smoothly with the latency. If we use the convention that undelivered messages due to, e.g., transmission errors, have infinite latency, then we can define the service reliability as the probability that the latency is less or equal to the deadline. That is, the service reliability can easily be found from the latency cumulative distribution function (CDF), see Figure 2 and Figure 3. Note that the end-to-end latency includes all delays that affect the message, e.g., MAC processing delays, channel access delays, transmission delays, retransmission delays, etc. Hence, it makes sense to model latency as a random quantity, which is completely characterized by its CDF.

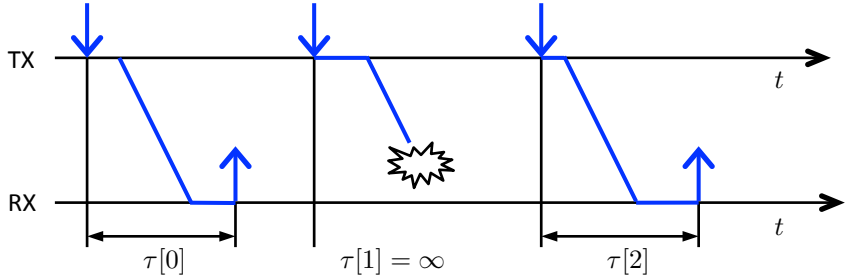

Figure 2: End-to-end latency realizations for three consecutive messages. The down arrows indicate when the sending-end application request transmission and the up arrows indicate when the receiving-end application receives error-free messages. By convention, undelivered messages have infinite latency.

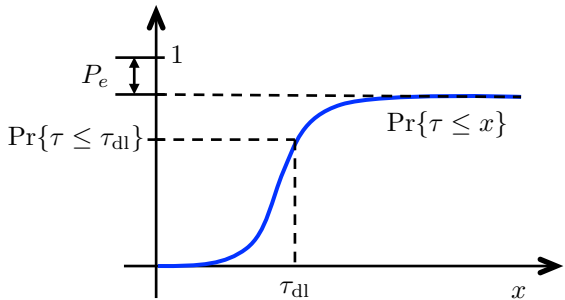

Figure 3: Latency CDF, reliability and message error probability. Reliability is found as the CDF evaluated at the deadline, and the probability of undelivered messages is found as the difference between 1 and the CDF asymptote.

Applications might also have requirements on jitter, i.e., the latency variation around its mean value. However, given that reliability is very high, we can reduce the jitter to an arbitrarily low value. The procedure is straightforward: we simply buffer messages at the receiving end and release messages from the buffer such that the overall latency (transmission latency plus buffer delay) is equal to the deadline. Since the transmission latency is upper limited (with a high probability) to the deadline, this procedure will force the overall latency to be essentially constant, i.e., to have very low jitter. Of course, the price we pay is in increased average latency, which might be problematic for applications with soft deadlines, but is of no consequence for hard deadline applications. Moreover, it is clear that we can modify the buffer release policy to reduce the average latency at the price of increased jitter. Hence, it seems that the main problem is to provide a service with high reliability, as this will give us tools to limit jitter. For this reason, we will not further discuss the problem of limiting jitter in this paper.

An ultra-reliable service typically provides very high reliabilities, e.g., 99.999% (5 nines). The deadline could be quite small, say on the order of milliseconds, but could also be more relaxed, e.g., on the order of seconds or higher. What is important is that, whatever the deadline is, an ultra-reliable service should deliver messages before the latency exceeds the deadline with very high probability.

It may be very costly or even impossible to provide ultra-reliable services at all times, due to unfavorable shadowing, excessive pathloss, intermittent high interference, etc. To make the uMTC service economically viable, it might therefore be needed to sometimes declare the uMTC service unavailable [Schotten-14]. That is, the application makes a request for service, which is either granted or not. If the service request is granted, then the communication system will transfer the message with the requested reliability. However, the communication system can also declare the service as unavailable, in which case the application needs to initiate a fallback procedure in order to limit the risk of accident, e.g., by reducing speed and increasing distances between platooning vehicles. Needless to say, the service availability, i.e., the probability that the uMTC service is declared as available, is an important performance metric for the communication system providing the service.

It is clear that we often have a tradeoff between availability and reliability, in the sense that we can make a system more reliable by reducing the availability, and vice versa. To illustrate this point, consider a highly simplified example of packet transmission over a fading channel (without retransmissions, rate control, or power control) when the transmitter has knowledge of the instantaneous signal-to-noise ratio (SNR). Furthermore, suppose reliability is bounded by the packet error probability (i.e., the latency CDF has reached its asymptotic value at the deadline, see Figure 3). We can then increase reliability by declaring the service as available only when the channel SNR is above a threshold. By increasing the SNR threshold, the transmitted packet error rate decreases and reliability increases. However, the availability also decreases, as the service will be declared as available more seldom as the SNR threshold increases.

We have now arrived at a point where we can describe a communication system that provides a uMTC service as a communication system that provides transmission of messages with very high reliability, whenever the system declares the service as available. When, on the other hand, the system declares the service as unavailable, the reliability of the messages cannot be guaranteed. Now, if the service is declared as available and if the requested reliability is not satisfied by the communication system, then we say that the uMTC service has failed. This event, i.e., uMTC service failure, has potentially catastrophic consequences and the probability for this must be made negligible. In other words, the system must make conservative estimates of its ability to satisfy the reliability requirements.

We conclude this section by offering formal definitions of uMTC service reliability, availability, and failure. Under the assumption that only error-free messages are delivered (i.e., perfect error detection is assumed) and that non-delivered messages have infinite delay, see Figure 1 and Figure 2, we define

- *Service reliability* as the probability that an arbitrary message is delivered error-free before a pre-described deadline
- *Service availability* as the probability that the service is declared as available
- *Service failure* as the probability that the service does not satisfy the pre-described reliability constraint, given that the service is declared as available

Indeed, we claim that service reliability, availability, and failure, are the main metrics that should be used when assessing the performance of ultra-reliable communication, such as safety-critical V2X communication. The basic notion of uMTC service reliability, availability, and failure will be further developed in the next section.

## 4 Reliable Service Composition and Availability Indication

The described event of failure of the uMTC service, described in the previous section, is based on the assumption that the application (the higher layer in Figure 1) puts forward a specific communication requirement on the communication service (the lower layer in Figure 1). This can be expressed in terms of a certain data amount/packet size that needs to be delivered within a certain deadline with a given probability. A generalization of this approach would be to have a stronger coupling between the two layers in Figure 1: the lower layer may send an indication to the higher layer about which type of service is available (e.g., maximal data rate and probability of delivery within a deadline), and the higher layer reacts by decreasing its communication requirements. This is the essence of the concept of Reliable Service Composition (RSC), which is a way to specify different versions of a service that is executed by the application layer, such that when the communication conditions are worsened, the Quality of Service (QoS) gracefully degrades to the service version that can be reliably supported. This type of operation replaces the traditional one based on binary decision "service available/not available" [Schotten-14]. The concept of graceful degradation of a service is not new, it has been used in, e.g., scalable video coding. However, video and its perception naturally allows for graceful degradation; in RSC, the objective is to design services that offer certain level of functionality when it is not possible to get the full one. Let us consider an example in the context of vehicle-to-vehicle communication [Popovski-14]. The basic version of the service is available 99.999% of the time. In the V2V setting, the basic version could involve transmission of a small set of warning/safety messages without certification. The fact that the set of messages transferable in the basic mode is limited can be used to design efficient low-rate mechanisms to transfer those messages. An enhanced version of the service is available 99% of the time and it includes limited certification and guarantees for transfer of payload of size D1 within time T1 with probability 99.9%. The full version is available 97% of the time, includes full certification and guarantees for transfer of payload of size D2 > D1 within time T2 < T1 with probability 99.9%.

In summary, the implementation of RSC relies on a careful consideration of the requirements set by the application and the availability indicator that the communication layer provides to the application layer.

## 5 Cooperative Intelligent Transport Systems

Cooperative intelligent transport systems (C-ITS) refers to a class of applications that rely on wireless communication to increase the safety and efficiency of the road transport system. The word "cooperative" signifies the underlying idea that road vehicles (cars, buses, trucks, two-wheelers, etc.) cooperate with each other and the road infrastructure (traffic signs, stop lights, road-side units, etc.) to avoid accidents, increase traffic flows, reduce travel time, reduce fuel consumption, and so on. Hence, vehicle-to-vehicle (V2V) and vehicle-to-road-infrastructure (V2I) communication are enablers of C-ITS applications. Recently, there has been an increased focus on enhancing safety for vulnerable road users (VRUs), such as cyclists and pedestrians. Hence, vehicle-to-device (V2D) communication is of great interest. In the following, we will collectively call all these communication modes for V2X communication.

C-ITS applications supported by ITS-G5 or DSRC rely mainly on two types of data traffic [Kenney-11, Strom-11]

- Periodic, time-triggered broadcast of status messages
- Event-triggered broadcast of warning messages

The messages include information about the vehicle position, heading, speed, etc. The messages are broadcasted to the vehicles in the immediate neighborhood, e.g., all vehicles inside a certain radius (the intended broadcast range) around the transmitting vehicle. This type of communication, i.e., when all vehicles are both transmitters and receivers of broadcasted messages is called *broadcast-broadcast* or *all-to-all broadcast,* to emphasize the multipoint-to-multipoint nature of the communication.

The state-of-the-art systems for V2X communication are called ITS-G5 in Europe and DSRC in the US [Strom-11, Kenney-11] and use dedicated spectrum near 5.9 GHz for traffic safety applications. Both systems use the PHY and MAC from the IEEE standard 802.11p. In fact, the 802.11p amendment is now classified as superseded and enrolled into the 2012 version of 802.11. Nevertheless, we will use the term "11p" when discussing the PHY and MAC used in ITS-G5 and DSRC. The only main innovation with 11p is that it allows for "communication outside the context of an BSS." This means that 11p nodes can start to communicate without first forming a basic service set (BSS), i.e., a regular Wi-Fi network. This is advantageous from a latency point of view, since V2X network topologies could be highly time-varying and forming and joining BSSs could incur significant delays. Interference is dominated by the intrasystem interference (due to the dedicated spectrum) and controlled by the 802.11 carrier-sense multiple access (CSMA) MAC.

V2X communication for C-ITS is characterized by highly dynamic network topologies, since communication nodes can move quickly in and out of radio range. This implies that a transmitter is unsure about how many receivers are within the intended broadcast range. Hence, it is cumbersome and potentially costly to implement retransmission (ARQ) protocols, and these are therefore not used in 802.11p-based networks. Moreover, since the wireless channel is also potentially highly time-varying, it is difficult, if not impossible, to provide strict reliability guarantees for 11p-based systems. This motivates to study whether 5G can replace or complement systems based on 11p. One advantage of 5G (or cellular in general) is that the basestation infrastructure can be used to control interference. 5G is also attractive since many VRUs are expected to carry 5G devices, which opens up for using V2D communication to enhance their safety. In principle, VRUs can alternatively be equipped with 11p-enabled devices, but at the moment, there seems to be no such devices on the market.

# 6  5G Technology Components to Support V2V According to TC12 Requirements

One of the METIS test cases, namely Test Case 12 (TC12), is concerned with ultra-reliable, low-latency communication for traffic safety and traffic efficiency applications [METIS D1.1]. The test case requirements are quite challenging: 1600 byte messages should be delivered within 5 ms with a reliability of 99.999%. Messages are sent either event-triggered or periodically time-triggered with a message rate up to 10 Hz, and the transmitter-receiver relative speed could be as high as 500 km/h. The availability is not specified in numbers, but should be very high [METIS D1.5].

In the following, we will describe a number of technology components that have been identified by METIS as promising to address the TC12 requirements. In short, we argue that it is suitable to use

- Device-to-device (D2D) links that are operational even as the network connectivity is limited or even non-existing;
- Radio resource management (RRM) based on slowly time-varying channel state information (CSI) or other slowly time-varying context information;
- Time-division multiplexing of transmissions, possibly generalized to time and frequency division;
- Channel estimation specifically tailored to V2X channels.

The choice of D2D links in favor of traditional uplink/downlink cellular links is motivated by the need to limit latency and to offer high availability. It is clear that every hop (uplink and downlink) consumes part of the latency budget, and it is therefore advantageous to use D2D, which only needs one hop. Moreover, no current cellular system has 100% coverage, and it seems unlikely that this will change in the future. There will likely be remote areas where there are roads, but too few xMBB customers to justify the cost incurred to provide 100% coverage. The coverage situation can vary from the extreme case when all uMTC nodes have network connectivity to the other extreme when no node is covered. Due to the high availability requirement, this implies that the D2D links must be operational also with limited or no network connectivity.

In the case when there is full or partial network connectivity, then network-assisted RRM (i.e., allocation of power, time, frequency, and possible other radio resources) is possible. However, due to the communication overhead and delay incurred by measuring and gathering CSI at the basestation, it is not reasonable to assume that the CSI is perfectly known by the RRM algorithm. Consequently, the two main RRM proposals from METIS rely only on slowly varying CSI, i.e., path-loss and large-scale (shadow) fading, or slowly varying context information in the form of vehicle positions. Details of the RRM approaches are found in [METIS D4.3, Sun-14, Sun-15, Botsov-14]. In both cases, uMTC users and xMBB users are assumed to share the spectrum, see Figure 4. To handle the sharing situation, a promising approach is to cast the RRM problem as an optimization problem with the uMTC service requirements as constraints and the xMBB service performance as the objective to be maximized. Hence, if the optimization problem is feasible, then the uMTC service requirements are guaranteed to be satisfied. Conversely, if the optimization problem is not feasible, this implies that the uMTC service is not available for some or all of the uMTC users. This RRM approach ties nicely into the RSC framework, which is explained in Section 4. Indeed, the availability metric is simply equal to the probability that the optimization problem is feasible.

October 3, 2015, Ver. 1.1

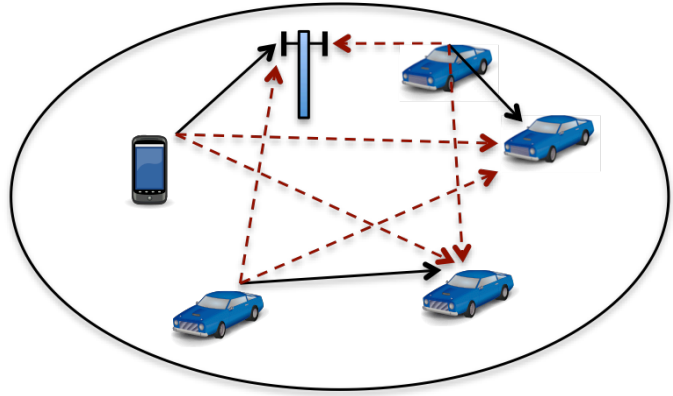

Figure 4: Example network with four uMTC users (vehicles) and one xMBB user (smartphone) that share the same uplink radio resource and are therefore interfering. The desired channels are marked with solid lines and interference channels with dashed lines.

Access to the wireless medium must be possible also in the extreme case when there is no network connectivity. Hence, we need to specify a MAC scheme for this case. The no network connectivity situation is very similar to the one faced by the 802.11p-based systems described in Section 5. One MAC option is therefore to use a CSMA-based approach, similar to the one in 802.11. However, to address the poor scaling behavior of CSMA and to control latency more precisely, we propose to use a time-division approach instead. Numerous distributed time-division multiple access (TDMA) schemes have been proposed in the literature. In particular, a scheme called self-organized TDMA (STDMA), which is used in the shipping industry, has been modified and proposed for C-ITS applications [Sjöberg-13]. STDMA is essentially a GPS-assisted dynamic reservation scheme that works well as long as the network topology does not change too rapidly. A more recent MAC approach, which is promoted by METIS, is based on coded slotted Aloha (CSA) and does not use reservations. The basic CSA scheme is modified to limit latency (by introducing a frame structure) and to support all-to-all broadcast [METIS D2.4, Ivanov-15]. It is assumed that each node needs to transmit one packet per frame and that the packet fits within one slot. Hence, the MAC-to-MAC delay is controlled by the frame duration. The main idea of CSA is that each node sends a random number of copies of its packet in uniform randomly chosen slots in the frame. Clearly, a slot can now contain zero, one, or multiple packets. A slot that contains exactly one packet is called a singleton slot. A receiver buffers the received frame and decodes all singleton slots (which we, for simplicity, assume can be done error-free). A decoded packet contains pointers to the other slots in which it is repeated. Hence, the receiver can attempt to cancel the contribution from the decoded packet in the whole frame. Successful cancelations might uncover new singleton slots, which can now be decoded. The decoding-cancellation process is repeated until no more singleton slots are uncovered. An example of the process is depicted in Figure 5. Simulation results presented in [Ivanov-15] show that CSA can support e.g. more than three times as many users at a reliability of 99.99%.

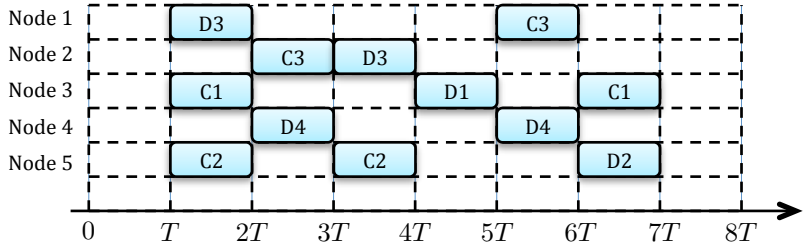

**Figure 5: CSA example with 8 slots per frame. Five nodes randomly select to transmit either two or three copies of their packets in random slots in the frame. Hence, the slots are populated with 0, 1, or more packets in a random fashion. By executing the CSA decode-and-cancel iteration, all packets can be decoded in this frame realization. In the first iteration, the packet marked with D1 is decoded and the copies marked C1 are canceled. A new singleton slot (slot 6) is revealed. Hence, packet D2 can be decoded and the packets marked with C2 are canceled. Continuing this process by decoding packet Dx and canceling the packets Cx for x = 3 and 4, we see that all packets are decodable.**

The CSA approach requires a minimum of coordination—in fact, only slot and frame synchronization is needed. Network synchronization can be derived from a global navigation satellite system (GNSS), such as GPS or Galileo, or from the cellular network infrastructure. However, just as cellular coverage is not 100%, neither is GNSS coverage. This situation, i.e., when connectivity to the synchronization infrastructure is limited or non-existing, can be addressed by applying network synchronization ideas developed in METIS, see [METIS D2.4] for details.

Communication with fast moving vehicles is challenging due to fast changing channel conditions. The time variations increase with the Doppler spread, which in turn increases with the carrier frequency and the speed with which vehicles and scatterers move. It is therefore important that we design channel estimation and channel prediction schemes that are able to handle highly time-varying channels.

Measurement campaigns have shown that the direct V2V channel exhibits significant structure in the delay-Doppler domain [Karedal-09], see Figure 6. Multipath components (MPCs) that result from interaction with large scatterers (stationary and moving) are relatively few and gives rise to strong sparse components in the delay-Doppler domain. The weaker diffuse MPCs are many more, but tend to occur in groups in the delay-Doppler plane. This mixed element-wise and group-wise sparsity can be exploited to enhance the channel estimation accuracy, with significant performance gains [Beygi-15].

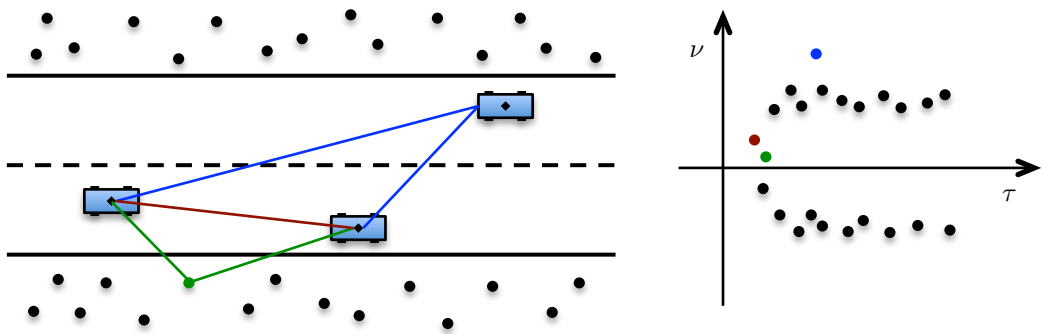

**Figure 6: Geometry of V2V propagation environment (left) and multipath components in the delay-Doppler domain ($\tau$-$\nu$). The scatterer in the geometry gives rise to MPCs. The line-of-sight path is indicated in red, a fixed scatterer MPC in green, and a moving scatterer MPC in blue.**

## 7 Conclusions and Outlook

The requirements for C-ITS applications are quite in line with the main purpose of the 5G uMTC service, namely to provide very reliable transfer of messages. Reliability is in this context defined as the probability that an arbitrary message is delivered to the receiving application within a pre-defined deadline. We have argued that a uMTC service for C-ITS applications should be composed by D2D links that are operational with varying degrees of network coverage. Hence, RRM and MAC should be flexible to leverage network coverage whenever possible and degrade gracefully as coverage is limited or non-existing. To make the uMTC service economically, or even physically, feasible, we argue that a RSC framework should be defined to support negotiation between the application using the uMTC service and the system that provides the service.

Although significant progress has been done during the lifetime of the METIS project, more work lies ahead before uMTC service for C-ITS applications becomes a practical reality. In particular, advances in RRM, MAC, and physical layer techniques for high-mobility D2D links are still required.

## 8 Acknowledgements

Part of this work has been performed in the framework of the FP7 project ICT-317669 METIS, which is partly funded by the European Union. The authors would like to acknowledge the contributions of their colleagues in METIS, although the views expressed are those of the authors and do not necessarily represent the project. In particular, we would like to acknowledge the contributions from Mladen Botsov, David Gozalvez Serrano, Peter Fertl, BMW, and Wanlu Sun, Mikhail Ivanov, Fredrik Brännström, and Alexandre Graell i Amat, Chalmers Univ. of Technology, and Raja Sattiraju, Univ. of Kaiserslautern.